\documentstyle[aps,twocolumn,epsfig]{revtex}
\begin{document}
\newcommand{\ve}[1]{\mbox{\boldmath $#1$}}
\twocolumn[\hsize\textwidth\columnwidth\hsize
\csname@twocolumnfalse%
\endcsname

\draft

\title{Auger decay, Spin-exchange, and their connection to 
Bose-Einstein condensation of excitons in Cu$_2$O} 
\author{G. M. Kavoulakis$^1$ and A. Mysyrowicz$^2$}
\date{\today} 
\address{$^1$NORDITA, Blegdamsvej 17, DK-2100 Copenhagen \O, Denmark
\\ $^2$Laboratoire d'Optique Appliqu\'ee, Ecole Nationale Sup\'erieure
de Techniques Avanc\'ees, \`Ecole Polytechnique, Palaiseau, France}

\maketitle

\begin{abstract}
   In view of the recent experiments of O'Hara {\it et al.}
\cite{Keith} on excitons in Cu$_2$O, we examine the interconversion 
between the angular-momentum triplet-state excitons and the angular-momentum
singlet-state excitons by a spin-exchange process which has been
overlooked in the past. We estimate the rate of this 
particle-conserving mechanism and find a substantially higher value 
than the Auger process considered so far. Based on this idea, we give a 
possible explanation of the recent experimental observations, and make 
certain predictions, with the most important being that the singlet-state 
excitons in Cu$_2$O is a very serious candidate for exhibiting the 
phenomenon of Bose-Einstein condensation.
\end{abstract}
 
\vskip0.5pc]

   Bose-Einstein condensation \cite{GSS} 
has been the subject of numerous theoretical and experimental studies.  
While in the recent years many of these studies have focused on the 
Bose-Einstein condensation of trapped alkali atoms \cite{RMP},
another possible candidate which can undergo this second-order phase 
transition is the exciton gas in semiconducting materials \cite{Knox}. 

  Excitons are much like positronium atoms. They are bound states 
which form between electrons and holes in a semiconductor,
after the electrons get excited from the conduction band to the
valence band, usually by some laser field.  Since excitons consist of 
two fermions, in the limit where their separation is much larger than 
their Bohr radius, they are expected to behave like bosons.  Many 
experiments have been performed with excitons in Cu$_2$O 
because of the many advantages of this material: It has a direct, but
dipole-forbidden gap, which makes the lifetime of excitons rather long,
it has isotropic effective electron and hole masses, it does not form 
bound states, biexcitons, or an electron-hole liquid, and finally the
exciton binding energy is quite large.

The traditional way of observing the kinetic energy distribution
of excitons is to look at the recombination spectrum, and more
specifically the optical-phonon assisted lines.  Since the optical 
phonons have a very weak dispersion relation and since the transition matrix
element does not depend on the exciton momentum, the energy 
distribution of the emitted photons essentially gives the kinetic 
energy distribution of the excitons.  Many experiments 
\cite{Andre,David} have demonstrated that excitons do indeed obey 
Bose-Einstein statistics in the limit of high enough densities and low 
enough temperatures, with the luminescence spectrum fitting very 
accurately to Bose-Einstein distributions.
This fitting procedure gives the temperature and the chemical 
potential of the gas, since these are essentially independent parameters.  
A crucial assumption underlying this procedure, is that very frequent
collisions between the excitons bring the gas to a quasi equilibrium, 
with some time-dependent chemical potential and temperature, which in 
general differs from the lattice temperature that is kept very low, 
below 5 K.  The typical effective temperature of the exciton gas is on 
the order of 10 up to 100 K.  Knowing the temperature and the chemical 
potential, one can deduce the particle density, assuming an ideal Bose 
gas with an experimentally known total exciton mass.  The densities turn 
out to be on the order of 10$^{18}$ cm$^{-3}$ from this method. Following 
this approach Snoke {\it et al.} \cite{Andre,David} observed that the
triplet-state (ortho)excitons do not Bose condense, but move along lines
parallel and closely to the critical one, which are adiabats, i.e.,
along lines with constant entropy per particle. This effect has been
examined theoretically in Ref.\,\cite{KBW}, and has been attributed
to a competition between the acoustic-phonon cooling of the exciton 
gas, and an Auger heating mechanism which prevents the Bose-Einstein
condensation of orthoexcitons. Lin and Wolfe have also reported in 
Ref.\,\cite{Jim} this tendency of orthoexcitons to move along adiabats, 
but, most importantly, have observed evidence for the Bose-Einstein 
condensation of the angular-momentum singlet state (para)excitons
\cite{KBW}.

However, recently O'Hara {\it et al.} have developed another method for 
estimating the density of the excitons \cite{Keith}.  By calibrating 
their photon detector, they have evaluated the number of photons that 
are being emitted, thus determining the number of orthoexcitons inside 
the crystal. They have also estimated the volume of the exciton gas by 
knowing the surface of the area of the laser light that creates the 
excitons, and with the assumption that the exciton gas has a typical 
depth inside the crystal, which is on the order of the absorption length 
of the laser light. Dividing the exciton number by the volume, they have 
found that the average density of the orthoexciton gas is two orders of 
magnitude smaller ($\sim 10^{16}$ cm$^{-3}$) than the one they estimate 
by the spectroscopic method. This implies that the gas should be completely
classical, without showing any kind of quantum degeneracy. 

Therefore one is confronted with a paradox, since from the one point of 
view the spectra can be fitted very accurately to Bose-Einstein 
distributions, but on the other hand the densities seem to be quite 
lower than those one gets by this fit, and certainly 
much lower than the region where one would observe Bose statistics.  
Furthermore, Refs.\,\cite{Keith,Snoke,KB} suggest that such low exciton 
densities might be due to a very effective Auger decay mechanism, 
where two excitons collide, the one recombines, transferring its 
energy to the other, which ionizes.  The implication is that this 
Auger mechanism, which does not conserve the total number of excitons 
prevents the onset of Bose-Einstein condensation.  Up to now the Auger 
process was thought to provide the only relatively fast channel for 
excitons to get destroyed \cite{KBW}.  However, in view of the very long 
intrinsic radiative lifetime of orthoexcitons reported recently in 
Refs.\,\cite{Keith}, the Auger decay rate should have a giant value, 
exceeding by three orders of magnitude the value calculated in 
Ref.\,\cite{KB}.

In this study we examine another mechanism \cite{otop},  
in which two orthoexcitons with opposite $J_z$ collide, where $\bf J$ 
is the total angular momentum of each exciton, exchanging their 
electrons or holes in the process, giving two paraexcitons in the
final state.  We make an estimate of the rate for this
spin-exchange process, and find that it is rather high. 
Based on the result of our calculation, we propose that this ortho-to-para
interconversion mechanism is actually the dominant one in 
the experimental conditions used so far. Strong experimental evidence
that our argument is true provide Fig.\,2 of the first paper 
in Refs.\cite{Keith}, Fig.\,5 of the second paper in Refs.\cite{Keith},
and Fig.\,4(a) of Ref.\,\cite{Jim}, where for late times it is clear 
that there is very slow decay of the paraexciton number. Consideration
of this process removes the contradiction described above between the two
methods that give the exciton density.  An important conclusion drawn 
from this new scenario is that the Bose-Einstein condensation of 
paraexcitons is probable, since at late times they should form a cold 
and relatively dense gas.  The spin-exchange mechanism, even if it 
converts one species into the other, conserves the total number 
of excitons.  Since the orthoexcitons lie higher in energy than the 
paraexcitons due to the exchange interaction by an amount $\Delta E$, 
the interconversion process also transfers energy to the exciton gas.
We also note that our mechanism explains the observed sublinear
dependence of the orthoexciton number \cite{Keith} that is generated by the
laser pulse as function of the laser power \cite{para}. On the other hand,
under extreme pumping conditions in the band-to-band region, we have a
highly nonequilibrium system at early times and one can think that Auger 
processes between the free carriers are also effective.

Let us start by making an order of magnitude estimate of the
decay rate $\Gamma_{o,p}$ of the ortho to paraexciton conversion process.  
We consider the process of two orthoexcitons with momenta $\bf K$ and
$\bf P$ and opposite $J_z$ colliding, giving two paraexcitons with momenta
${\bf K}'$ and ${\bf P}'$. Fermi's golden rule gives for the rate 
\begin{eqnarray}
   \Gamma_{o,p} = \frac {2 \pi} {\hbar} 
  \sum_{{\bf K},{\bf P},{\bf K}',{\bf P}'} |M|^2
 f_{\bf K}^o f_{\bf P}^o (1+f_{{\bf K}'}^p) (1+f_{{\bf P}'}^p)
\nonumber \\ \times
\delta(E_{{\bf K}'} + E_{{\bf P}'} - E_{\bf K} - E_{\bf P} -2
\Delta E)
\, \delta_{{\bf K}+{\bf P},{\bf K}'+{\bf P}'},
\label{gr}
\end{eqnarray}
where $M$ is the matrix element for this process, 
$f_{\bf K}^i$ is the distribution function of the $i$ species
(ortho or para excitons), having a dispersion relation $E_{\bf K}=
\hbar^2 K^2/2m$, with $m$ being the total exciton mass.
In this crude calculation we consider a cold (${\bf K}, {\bf P} 
\approx 0$) orthoexciton gas, which allows us to 
write that
\begin{eqnarray}
   \Gamma_{o,p} \approx \frac {2 \pi} {\hbar} N_o^2 \, \sum_{{\bf 
   K}',{\bf P}'}  |M|^2
 (1+f_{{\bf K}'}^p) (1+f_{{\bf P}'}^p)
\nonumber \\ \times \delta(E_{{\bf K}'}^p + E_{{\bf P}'}^p - 2 \Delta E)
 \, \delta_{{\bf K}'+{\bf P}',0},
\label{grr}
\end{eqnarray}
where $N_i$ is the total number of excitons of species $i$.
The energy conservation condition in the above equation
implies that $K'$ and $P'$ are of order $(m \Delta E/\hbar^2)^{1/2}$. 
Since for these wavevectors
the occupation number is much less than 1, we can ignore the 
enhancement factors $1+f$ above.  In addition, we argue below that the 
typical momentum exchange that enters the matrix element $M$
is of order $a_B^{-1} \sim (m E_b/\hbar^2)^{1/2} \gg 
(m \Delta E/\hbar^2)^{1/2}$, where $a_B $ is the exciton Bohr radius,
and $E_b$ is the exciton binding energy.
Since $E_b \gg \Delta E$, $M$ does not vary substantially in the sum 
of Eq.\,(\ref{grr}) and can be taken outside it, 
\begin{eqnarray}
   \Gamma_{o,p} \approx \frac {2 \pi} {\hbar} \frac {N_o^2} 2 \, |M|^2
   \sum_{{\bf K}'} 
\delta(E_{{\bf K}'}^p -  \Delta E).
\label{grrr}
\end{eqnarray}
The last sum is simply the density of states calculated for an energy 
$\Delta E$. The interaction that enters the matrix element $M$
is some screened Coulomb potential $V(q,\omega)$,
\begin{eqnarray}
  V(q,\omega) = \frac {4 \pi e^2} {\Omega \epsilon({\bf q},\omega) q^2}.
\label{me}
\end{eqnarray}
where $\Omega$ is the volume of the crystal, and $\epsilon({\bf q},\omega)$
is the dielectric function \cite{KCB}. 
The exciton wavefunction $\Phi_{\bf K}({\bf r}_e-
{\bf r}_h)$ of an exciton carrying momentum $\bf K$ can be written as
\begin{eqnarray}
   \Psi_{\bf K}({\bf r}_e-{\bf r}_h) = \frac 1 {\sqrt \Omega} \,
  e^{i {\bf K} \cdot ({\bf r}_e+{\bf r}_h)/2} 
 \sum_{\bf q} \phi_{\bf q} \, e^{i {\bf q} \cdot ({\bf r}_e-{\bf r}_h)},
\label{excwf}
\end{eqnarray}
where ${\bf r}_e$ and ${\bf r}_h$ are the electron and hole coordinates, 
and $\phi_{\bf q} = 8 (\pi a_B^3)^{1/2}/[1+(q a_B)^2]^2$ is the Fourier 
transform of the ground state hydrogenic wavefunction, $\Phi = e^{-r/a_B}/
(\pi a_B^3)$, which we assume to be the relative electron-hole 
wavefunction. We have assumed that the two colliding orthoexcitons
have ${\bf K}={\bf P}=0$; denoting the momenta of the electron and the hole
in each pair as ${\bf k}, -{\bf k}$, and ${\bf p}, -{\bf p}$ [see Fig.\,1], 
after the two excitons have exchanged their electrons or holes, and some 
momentum $\bf q$, conservation of energy and momentum requires that
\begin{eqnarray}
  \frac {\hbar^2 ({\bf k}-{\bf p}+{\bf q})^2}  {2m} = \Delta E.
\label{conservation}
\end{eqnarray}
\begin{figure}
\begin{center}
\epsfig{file=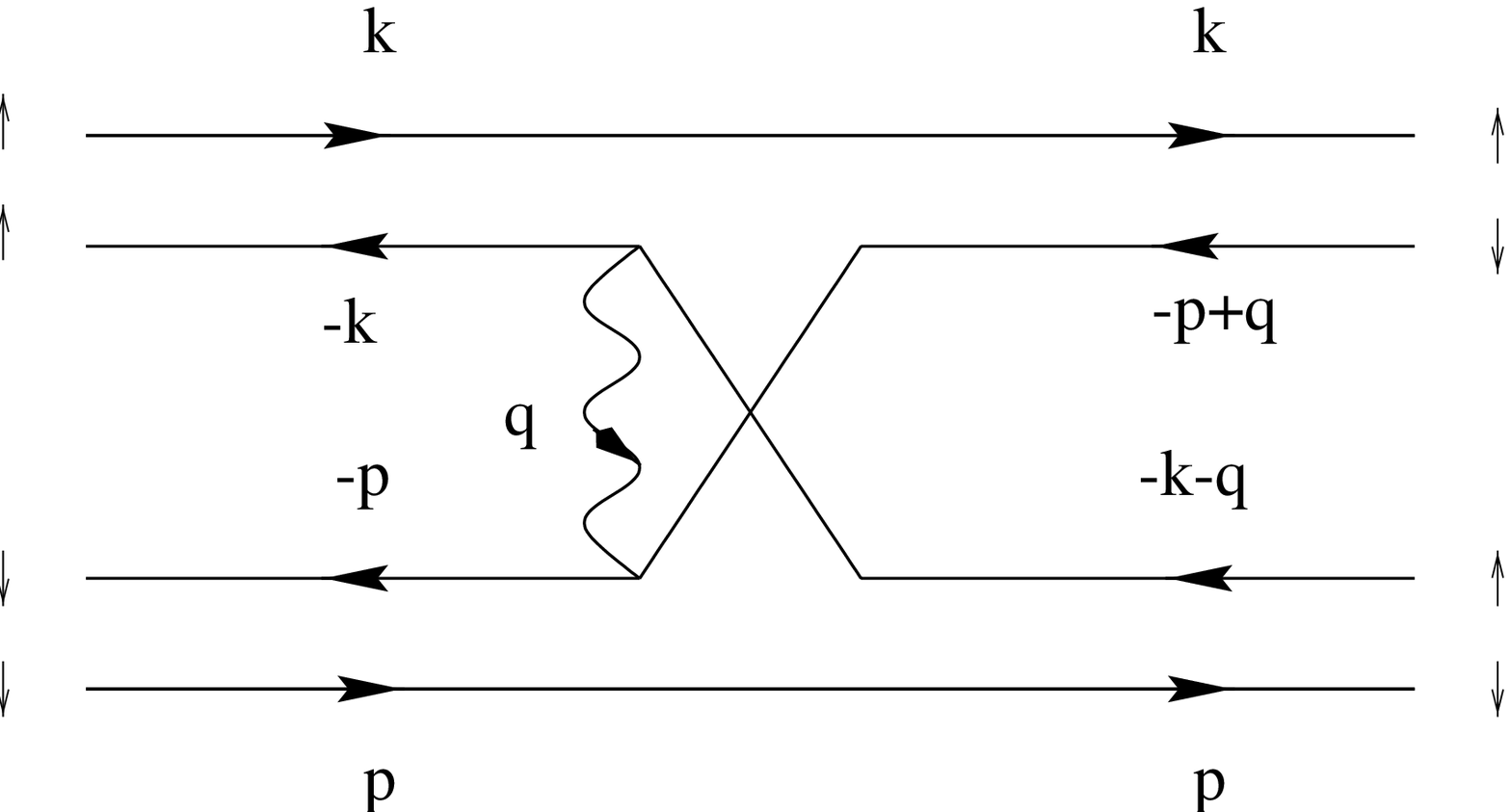,width=6.0cm,height=4.0cm,angle=0}
\vskip1pc
\begin{caption}
{The process of two orthoexcitons with opposite $J_z$ colliding, 
exchanging their holes and some momentum $\bf q$, giving two
paraexcitons in the final state. Time progresses from left to right.
The straight lines with arrows pointing to the right (left) denote
electrons (holes). The wiggly line denotes the Coulomb interaction.
The arrows pointing upwards (downwards) denote $J_z=1/2$ ($J_z=-1/2$)
electrons or holes.}
\end{caption}
\end{center}
\label{FIG1}
\end{figure}
\noindent
Since the exciton wavefunction has a momentum spread of order $a_B^{-1}$,
we see that $p \sim k \sim a_B^{-1}$, or in other words, $\hbar^2 p^2/2m
\sim E_b$. Because of the condition $\Delta E \ll E_b$, we conclude that
$q$ is also of order $a_B^{-1}$ \cite{det}. Furthermore, it is a rather 
good approximation 
to assume that $\epsilon({\bf q},\omega) \approx \epsilon_0$ \cite{KCB},
the low-frequency dielectric constant of Cu$_2$O, which is approximately
equal to 7.5, and thus
\begin{eqnarray}
  |M| \approx \frac {4 \pi e^2 a_B^2} {\Omega \epsilon_0}.
\label{mee}
\end{eqnarray}
After some trivial manipulations we can write for the rate of the 
orthoexciton conversion process $\Gamma_{o,p}/\Omega = n_o/
\tau_{o,p}$, where $n_i$ is the density of species $i$, and
\begin{eqnarray}
    \tau_{o,p}^{-1} \approx
  16 \pi \, n_o a_B^3 \, \sqrt \frac {\Delta E} {E_b} \,
 \frac {E_b} \hbar.
\label{rate}
\end{eqnarray}
Let us now examine the reverse process of two paraexcitons colliding,
giving two orthoexcitons. The energy scale which is very important for
this process is the energy splitting $\Delta E$ between the orthoexcitons
and paraexcitons in Cu$_2$O, which is equal to 
12 meV \cite{KCB} at the zone center, corresponding to a temperature
of approximately 150 K. We can write for the decay rate
$\Gamma_{p,o}/\Omega = n_p/\tau_{p,o}$, with
\begin{eqnarray}
  \tau_{p,o}^{-1} 
   \approx \tau_{o,p}^{-1} \, (n_p/n_o) \, e^{-\Delta E/ k_{B} T},
\label{reverse}
\end{eqnarray}
where $T$ is the temperature of the exciton gas. Thus, for temperatures
which are much lower than $\Delta E/k_{B}$, the interconversion
mechanism converts the orthoexcitons into paraexcitons, but not the
reverse, since the para-ortho process is exponentially suppressed 
$[e^{-\Delta E/ k_{B} T} \approx 0]$.  On the other hand, for 
temperatures which are $\agt \Delta E/k_{B}$, the two rates $\Gamma_{o,p}$ 
and $\Gamma_{p,o}$ can be comparable, depending on the relative densities 
of ortho and paraexcitons, and thus the net rate can be very low.

To calculate the actual value of the rate given by Eq.\,(\ref{rate}), we
use the following numbers for excitons in Cu$_{2}$O (which have 
very low uncertainty): the binding energy is 153 meV, the energy 
splitting $\Delta E$ at the zone center is 12 meV, and finally the Bohr 
radius $a_{B}$ is 5.3 \AA, as given by a variational calculation 
presented in Ref.\,\cite{KCB}.  With these numbers we get
\begin{eqnarray}
   \tau_{o,p}^{-1} \approx 5 n_o(10^{16} \rm{cm}^{-3}) \, \rm{ns}^{-1},
\label{numrr}
\end{eqnarray}
where the notation $n(10^{16} \rm{cm}^{-3})$ means that the density
is to be measured in units of $10^{16} \rm{cm}^{-3}$. 

We can now compare the experimental decay rate of orthoexcitons that 
was measured recently by O'Hara {\it et al.} \cite{Keith} with the
above theoretical number on the one hand, and with the theoretical
number for the Auger process on the other hand.
For the low lattice temperature of $\sim 2$ K in this experiment, the 
average kinetic energy per particle of the gas of orthoexcitons is 
expected to be much less than $\Delta E$, and thus we assume
for the net interconversion rate $\Gamma_{o,p} - \Gamma_{p,o} \approx
\Gamma_{o,p}$. The ``two-body decay constant'', $A \approx 10^{-16}$ 
cm$^3$/ns, that is extracted from the data given in Eq.\,(5) of the first
paper in Refs.\,\cite{Keith} has the same order of magnitude as the
rate given by Eq.\,(\ref{numrr}).

We mentioned before that the Auger process is expected theoretically 
to have a rather low decay rate, in view of the very long radiative
lifetime of orthoexcitons which was reported recently in Refs.\,\cite{Keith}.
More specifically, in the theoretical study of the Auger process 
described in Ref.\,\cite{KB}, a detailed analysis of this mechanism
implied that the phonon-assisted Auger decay process was the dominant
one.  However, to get the rate, the authors used the orthoexciton 
radiative lifetime of 25 ns at a temperature of 10 K \cite{lifet} that 
was measured in the past.  In Refs.\,\cite{Keith} the same quantity was 
measured to be approximately 10 $\mu$s, and since the radiative lifetime 
can be limited by imperfections or any other factor, the real radiative 
orthoexciton lifetime is at least 10 $\mu$s, or even longer.  This 
implies that the phonon-assisted Auger decay rate, based on the 
theoretical study of Ref.\,\cite{KB} is negligible.  These arguments 
provide strong evidence that the ortho to para exciton interconversion 
mechanism that we study here is really the dominant process.

We turn now to the contradiction between the two methods which 
have been used for determining the orthoexciton density.
It is important to get first an estimate for the interparticle elastic
scattering rates, in order to compare them with the rates 
of interconversion by spin-exchange.  Typical scattering rates 
$\tau^{-1}$ for elastic collisions between excitons are expected to be 
on the order of  
$\tau^{-1} = n \sigma v_{\rm th},$
where $\sigma$ is the scattering cross section, and $v_{\rm th}$ is
the thermal velocity, $v_{\rm th} = (8 k_{B} T/\pi m)^{1/2}$.  At the low 
temperatures of interest one can assume hard-sphere scattering between
the excitons. If $a$ is the scattering length, then for identical bosons
$\sigma = 8 \pi a^{2}$.  Recently the scattering length for excitons in
Cu$_{2}$O has been calculated to be on the order of $2 a_{B}$ \cite{Ceperley}
with use of Monte-Carlo simulations.  Therefore we get 
\begin{eqnarray} 
   \tau^{-1} \approx 0.1 \, n(10^{16} \rm{cm}^{-3}) \,
   \sqrt {T (\rm{K})} \, \rm{ns}^{-1},
\label{scatn}
\end{eqnarray}
assuming that the total exciton mass is equal to 3 electron masses
\cite{KCB}.  Here the density is measured in units of $10^{16}$ 
cm$^{-3}$, and the temperature in units of degrees Kelvin.  

After the laser pulse starts to decrease, the paraexcitons become the
dominant component of the gas because of the interconversion 
mechanism. For typical paraexciton densities of 
$10^{17}$ cm$^{-3}$ and thermal velocities of order 50 K, 
one sees from Eq.\,(\ref{scatn}) that the typical scattering times are 
of order 100 ps, and the paraexcitons should be able to
establish thermal equilibrium.  The paraexcitons should also have a 
well-defined chemical potential, since after the initial times that 
the orthoexcitons convert to paraexcitons due to the interconversion 
process, their number does not vary significantly with time. By contrast,
the orthoexciton-orthoexciton elastic scattering processes become
less and less frequent because of their decreasing density, even if the 
orthoexciton-paraexciton collisions can bring them to thermal 
equilibrium; we claim, however, that chemical equilibrium has
not been established in the orthoexciton gas, since the orthoexcitons
have a relatively fast way of converting into paraexcitons and their 
number is not conserved. One can speculate that under such circumstances 
the orthoexciton gas could have a chemical potential which is rather low, 
but this is a non-equilibrium problem, and it requires a detailed study.
One, for example, could use the Boltzmann equation to describe all the 
important processes which take place, and derive the distribution 
function of the orthoexcitons as a function of time.

There are some remarks one can make concerning this model we
are proposing. Firstly, the estimate we made for the interconversion rate, 
Eqs.\,(\ref{gr}) and (\ref{grr}), does not assume thermal equilibrium, 
which is rather important in our problem.  Secondly the Bose-Einstein 
condensation of paraexcitons does not seem hard, since as we explained 
there is no efficient mechanism which would destroy them on the 
timescales of interest, and their expansion could be the only factor 
against them in order to Bose condense.  However the expansion is not 
expected to be dramatic, and it can also be reduced by applying some 
stress to the crystal, thus effectively trapping the excitons.

  Finally, it has been argued is Ref.\,\cite{KBW}, that the 
orthoexcitons -- provided they are not far away from equilibrium --
have to move along lines of constant entropy, along which $n_{o} \propto
T^{3/2}$. To derive this result, the authors assumed that there is a 
competition between acoustic-phonon cooling and Auger heating. Since the phonon
cooling rate is $\propto -T^{3/2}$ for low lattice temperatures, and the Auger
heating rate is $\propto n_{o}$, $n_{o} \propto T^{3/2}$. Remarkably,
if indeed the Auger process is negligible, and the interconversion process
is the dominant mechanism, the heating rate due to this effect is equal to
$\Delta E /\tau_{o,p}$, and thus still proportional to $n_{o}$. Since this 
argument does not depend on the quantum degeneracy of the gas, $n_{o}$ should
still be proportional to $T^{3/2}$: we conclude again that the orthoexcitons
are expected to move parallel to the phase boundary, along adiabats, in 
contrast to the paraexcitons which most probably Bose condense.  
More experimental and theoretical work are required to verify these 
predictions. 

\vskip0.5pc

G.M.K. was supported by the European Commission, TMR program, contract
No.\,ERBFMBICT 983142.  Helpful discussions with K. Johnsen are
gratefully acknowledged.  
G.M.K.  would like to thank the Foundation of Research 
and Technology, Hellas (FORTH) for its hospitality. 
A.M. is grateful to the Humbolt foundation for support during this work.


\begin{references}
\bibitem{otop} This process has to be distinguished
   from the phonon-assisted ortho-to-para exciton interconversion
   mechanism, a process which has already been explored in Ref.\cite{op}
   (see also Ref.\,\cite{KBW}) and which does not depend on the particle 
   density. We will not consider it further here, but it would certainly have 
   to be taken into account in a complete quantitative study of the problem.
\bibitem{para} We should stress out that the paraexcitons
   have only  a very weak phonon-assisted line in the recombination
   spectrum (their direct recombination is highly forbidden \cite{KCB}),
   which makes their observation impossible at early times, during the
   action of the laser pulse.
\end{references}
\end{document}